\documentclass{article}
\usepackage{amsfonts,amssymb, amsmath}
\usepackage[english]{babel}

\DeclareMathAlphabet{\mathpzc}{OT1}{pzc}{m}{it}

\def\nn{\nonumber }
\def\bq{ \begin{equation} }
\def\eq{ \end{equation} }
\def\ben{ \begin{eqnarray} }
\def\en{ \end{eqnarray} }

\newtheorem{re}{Remark}
\newenvironment{rem}{\begin{re} \rm }{\end{re}}
\begin{document}


\title{On bi-hamiltonian geometry of the  Lagrange top}
\author{ A V Tsiganov \\
\it\small
St.Petersburg State University, St.Petersburg, Russia\\
\it\small e--mail: tsiganov@mph.phys.spbu.ru}

\date{}
\maketitle

\begin{abstract}
We consider three different incompatible bi-Hamiltonian structures for the  Lagrange top, which have the same foliation by symplectic leaves. These bivectors may be associated with the different 2-coboundaries in the Poisson-Lichnerowicz cohomology defined by canonical bivector  on $e^*(3)$.
\end{abstract}

\par\noindent
PACS: 45.10.Na, 45.40.Cc
\par\noindent
MSC: 70H20; 70H06; 37K10

\vskip0.1truecm

\section{Introduction}
\setcounter{equation}{0}
In recent years considerable progress has been made in investigations of the
bi-in\-teg\-rab\-le systems with functionally independent integrals of motion
\[
\{H_i,H_j\}=\{H_i,H_j\}'=0,\qquad  i,j=1,\ldots,n,
\]
where $\{.,.\}$ and $\{.,.\}'$ are compatible Poisson brackets on the
bi-hamiltonian manifold $M$.

Historically the majority of them come from stationary flows, restricted flows or the Lax equations of underlying soliton systems (see references in \cite{bl98,fp02,mag97} ). Construction of integrals of motion for such systems is usually based on the Lenard-Magri recurrence relations. In order to solve the corresponding equations of motion in framework of the separation of variables method we have to use some suitable reductions of the Poisson bivectors \cite{fp02,mt02}.

The other class of bi-integrable systems come from $r$-matrix algebras,  classifications of 2-coboundaries in the Poisson-Lichnerowicz cohomology and the separation of variables method \cite{compf95,ts07a,ts07b,ts07c}. The corresponding Poisson brackets $\{.,.\}$ and $\{.,.\}'$ have a common foliation as their symplectic leaf foliations. In this case we lose benefits given by bi-hamiltonian recurrence relations, but we can obtain the separated variables directly.

The main aim of this note is to discuss different bi-hamiltonian structures of both types for the Lagrange top.

\section{The bi-hamiltonian manifolds}
In this section we describe the manifolds where our bi-integrable systems will be defined.

Let $ M$ be a finite-dimensional Poisson manifold endowed with a  bivector $P$ fulfilling the Jacobi condition
\[
[P,P]=0
\]
with respect to the Schouten bracket $[.,.]$. We will suppose that $P$ has the constant corank $k$,  dim$M=2n+k$, and  that $C_1,\ldots,C_k$ are globally defined independent Casimir functions on $ M$
\[PdC_a=0,\qquad a=1,\ldots,k.\]
The $2n$ dimensional symplectic leaves of $P$ form a symplectic foliation.

A bi-Hamiltonian manifold $M$ is a smooth  (or complex) manifold endowed with two compatible bivectors $P,P'$ such that
\bq\label{m-eq}
[P,P]=[P,P']=[P',P']=0.
\eq
Classification of compatible Poisson bivectors  on low-dimensional Poisson manifolds is nowadays a subject of  intense research. However the higher dimensional problem is virtually untouched.

\subsection{Integrals of motion from the Poisson bivectors}

Let us consider bi-Hamiltonian manifold $M$ with some known bivectors $P$ and $P'$.
Moreover, let as suppose that there are $k$ polynomial Casimir
functions of the Poisson pencil $P_\lambda =P'-\lambda  P$,
\begin{equation}
\label{polcas}
H^{a}(\lambda )=\sum_{i=0}^{n_a} H^{a}_{i}\lambda^{n_a-i}\ ,\qquad H^{a}_0=C_a\,,\qquad
a=1,\dots,k\
,
\end{equation}
such that $n_1+n_2+\cdots+n_k=n$ and such that the
differentials of the coefficients $H^{a}_i$ are linearly independent on
$M$.

The collection of the $n$ bi-hamiltonian vector fields
\begin{equation}
\label{eq:01}
X^{(a)}_i=P\, dH^{a}_{i}=P'\,dH^{a}_{i-1}\ ,\quad
i=1,\dots,n_a,\quad
k=1,\dots,a\ ,
\end{equation}
associated with the Lenard-Magri sequences defined by the Casimirs
$H^{a}(\lambda)$ is called the { Gel'fand--Zakharevich  system}.

The standard arguments from the theory of Lenard--Magri chains show
that all the coefficients $H^{a}_{i}$ (\ref{polcas}) pairwise commute with respect to both
$\{.,.\}$ and $\{.,.\}'$. It allows us to get nontrivial bi-integrable systems with
 integrals of motion $H^a_i$ starting from the Casimir functions $H^{a}_0=C_a$ only.

If there exists a foliation of $M$, transversal to the symplectic leaves of $P$ and compatible with the Poisson pencil (in a suitable sense), then the restrictions of the
Gel'fand--Zakharevich systems on symplectic leaves of $P$ are separable in the so-called Darboux-Nijenhuis variables \cite{fp02}.

Summing up, if we have two compatible Poisson bivectors $P, P'$ and the Casimir functions $C_a$ of $P$, then we can get integrals of motion $H_i$ in the bi-involution using recurrence relations (\ref{eq:01}) and, if we are lucky, then we obtain the separated variables after some appropriate reduction.

\subsection{The Poisson bivectors from integrals of motion }
Let us consider bi-Hamiltonian manifold $M$ with canonical bivector $P$ and some integrable system with integrals of motion $H_m$. According to the Liouville-Arnold theorem any integrable system admits separation of variables in the action-angle coordinates.

According to \cite{skl95} integrable system on the symplectic leaves of $ M$   will be said to be separable in a set of canonical variables $(p,q)=(p_1,\ldots,p_n,q_1,\ldots,q_n)$ if there exist $n$ separated equations of the form
\[
\phi_j(p_j,q_j,\alpha_1,\ldots,\alpha_n, C_1,\ldots,C_k)=0, \qquad \mbox{\rm det}\left[\dfrac{\partial \phi_i}{\partial \alpha_j}\right]\neq 0,\qquad \{q_i,p_i\}=1.
\]
If we resolve these equations  with respect to parameters $\alpha_1,\ldots, \alpha_n$ one gets $n$ independent integrals of motion
\bq\label{m-int}
\alpha_m=H_m(p,q,C),\qquad m=1,\ldots,n,
\eq
as functions on the phase space $ M$ with coordinates $z=(p,q,C)$.

These integrals of motion $H_i(p,q,C)$ are in the involution
 \bq\label{inv-f}
 \{H_i,H_j\}_f=0,\qquad i,j=1,\ldots,n,
 \eq
 with respect to the following bracket $\{.,.\}_f$ on $M$
\ben
\label{poi-f}
\{q_i,p_j\}_f&=&\delta_{ij}\,f_j(p_j,q_j)\,,\\
\{p_i,p_j\}_f&=&\{q_i,q_j\}_f=\{p_i,C_j\}_f=\{q_i,C_j\}_f=\{C_i,C_j\}_f=0,\nn
\en
which depends on $n$ arbitrary functions $f_1(p_1,q_1),\ldots,f_n(p_n,q_n)$ \cite{ts07}.
This bracket defines  the  Poisson bivector
\bq
\label{f-ten}
P^f=\left(
 \begin{array}{ccc}
 0 & \mbox{\rm diag}(f_1,\ldots,f_n)&0 \\
 -\mbox{\rm diag}(f_1,\ldots,f_n) & 0&0\\
 0&0&0
 \end{array}
 \right)\,,
\eq
compatible with the canonical Poisson bivector $P$ on $M$ and such that
\bq\label{m-eq3}
P^fdC_{a}=0,\qquad a=1,\ldots,k.
\eq
If all the $f_i\neq 0$, then $P^f$ has the same foliations by symplectic leaves as  $P$. So, it is explicit construction of the Poisson structures having the same foliation by symplectic leaves.

Compatibility conditions (\ref{m-eq}),  equations (\ref{inv-f}) and (\ref{m-eq3}) may be checked in any coordinate system  on $M$. So, for any integrable system on $M$  we can try to solve the following system of equations
\bq\label{m-sys}
[P,P^f]=[P^f,P^f]=0,\qquad P^fdC_{a}=0,\qquad
\{H_m,H_l\}_f=0
\eq
with respect to $P^f$. Obviously enough, in their full generality  equations (\ref{m-sys}) are too difficult to be solved because it has infinitely many solutions labeled by  different separated coordinates and their functions $f=(f_1,\ldots,f_n)$ \cite{ts07}. In order to get particular solution of the system (\ref{m-sys}) we have to use some addition assumptions or the couple of ans\"{a}tze.

Nevertheless,  using any known solution $P^f$ of (\ref{m-sys}),  we can easily to solve the following system of algebraic equations
\bq\label{f-eq}
P^fdH_m=\sum_{l=1}^n F_{ml}dH_l,\qquad m=1..n,
\eq
with respect to entries of the $n\times n$ control matrix $F$. According to \cite{fp02} the eigenvalues of the control matrix $F$ are  the separated coordinates
\[
\det(F-\lambda \mbox{I})=\prod_{j=1}^n(\lambda-q_j)\,.
\]
Solution of the equations (\ref{m-sys}) -- (\ref{f-eq}) may be considered as a direct method of computation of the separated coordinates $q_j$ starting with given  integrals of motion only.

\begin{rem}
In fact  we postulate in (\ref{m-int}) and (\ref{poi-f}) that our separated variables are "invariant" with respect to the Casimirs, as the one considered in \cite{ts01}.
\end{rem}

\begin{rem}
Bivectors $P'$ fulfilling the compatibility condition $[P,P']=0$ are called 2-cocycles in the Poisson-Lichnerowicz cohomology defined by $P$ on $M$ \cite{lih77}.  The Lie derivative of $P$ along any vector field $X$ on $M$
\bq\label{co-b}
P'=\mathcal L_X(P)
\eq
is 2-coboundary, i.e. it is 2-cocycle associated with the Liouville vector field  $X$. For such bivectors   $P'$ the compatibility conditions (\ref{m-eq}) are reduced to the single  equation
\bq\label{m-eq2}
[\mathcal L_X(P_0),\mathcal L_X(P_0)]=0,\quad\Leftrightarrow\quad\
[\mathcal L_X^2(P_0),P_0]=0
\eq
The second Poisson-Lichnerowicz cohomology group $H^2_{P_0}(M)$ of $M$ is precisely the set of bivectors $P_1$ solving $[P_0,P_1]=0$ modulo the solutions of the form $P_1=\mathcal L_X(P_0)$. We can interpret  $H^2_{P_0}(M)$ as the space of infinitesimal deformations of the Poisson structure modulo trivial deformations.  For regular Poisson manifolds cohomology reflect the topology of the leaf space and the variation in the symplectic structure as one passes from one leaf to another.

In our case  the components of the Liouville vector field  $X$ in the variables $(p,q,C)$ are equal to
\[
X_j=\left\{\begin{array}{cl}F_j(q_j,p_j), \qquad& j=1,\ldots,n\\
0,\qquad&j=n+1\ldots 2n+k\end{array}\right.
\]
and bivector $P^f=\mathcal L_X(P)$ has the form (\ref{f-ten}) with
\[
f_j(q_j,p_j)=-\frac{\partial}{\partial q_j}\, F_j(q_j,p_j)\,.
\]

So, in fact in the separation of variables method  we are looking for special 2-coboundaries (\ref{f-ten}) having the same foliation by symplectic leaves as canonical bivector $P$ (\ref{m-eq3}).
\end{rem}

Summing up, if we have the canonical Poisson bivectors $P$, it's Casimir functions $C_a$ and integrals of motion $H_m$ for some integrable system, then we can try to get compatible  Poisson bivector $P^f$ from the equations (\ref{m-eq}), which immediately gives rise to the corresponding separated variables.

\section{The Lagrange top}
\setcounter{equation}{0}

Let two vectors $J=(J_1,J_2,J_3)$ and $ x=(x_1,x_2,x_3)$ are coordinates on
the Euclidean algebra $e(3)^*$ with the Lie-Poisson bracket
\begin{equation}\label{e3}
\,\qquad \bigl\{J_i\,,J_j\,\bigr\}=\varepsilon_{ijk}J_k\,, \qquad
\bigl\{J_i\,,x_j\,\bigr\}=\varepsilon_{ijk}x_k \,, \qquad
\bigl\{x_i\,,x_j\,\bigr\}=0\,,
\end{equation}
where $\varepsilon_{ijk}$ is the totally skew-symmetric tensor.
This bracket has two Casimir functions
\bq \label{caz-e3}
C_1=|x|^2\equiv\sum_{k=1}^3 x_k^2, \qquad C_2= (x,J)\equiv\sum_{k=1}^3 x_kJ_k .
\eq
Fixing their values one gets a generic symplectic leaf of $e(3)$
\[
{\mathcal O}_{ab}:\qquad \{{x}\,, {J}\,:~C_1=\alpha^2,~~
C_2=\beta\}\,,
\]
which is a four-dimensional symplectic manifold.
 As usual we identify
$(\,\mathbb R^3,\wedge)$ and  $(so(3),[.,.])$ by using the well known isomorphism of the Lie algebras
\bq\label{trans-M}
 z=\left(z_1,z_2,z_3\right)\to z_M=\left(\begin{smallmatrix}
            0 & z_3 & -z_2 \\
            -z_3 & 0 & z_1 \\
            z_2 & -z_1 & 0
          \end{smallmatrix}\right),
\eq
where $\wedge$ is the cross product in $\mathbb R^3$ and  $[.,.]$ is the matrix commutator in $so(3)$. In these coordinates the canonical Poisson bivector on  $e^*(3)$
is equal to
\bq \label{p0-e3}
 P=\left(\begin{array}{cc}0&x_M\\x_M&
J_M\end{array}\right).
\eq

The Lagrange top is one  of the most classical examples of integrable systems
with the following integrals of motion
\bq\label{int-lag}
H_1=J_3,\qquad H_2=J_1^2+J_2^2+J_3^2+ax_3,\qquad a\in \mathbb R.
\eq
Here $J$ and $x$ denote respectively the angular momentum and the coordinates of the unit vector in the direction of gravity, all expressed in the body frame.

This is a special case of rotation of a rigid body around a
fixed point in a homogeneous gravitational field, characterized by the following conditions: the rigid body is rotationally symmetric, i.e. two of its three principal moments of
inertia coincide, and the fixed point lies on the axis of rotational symmetry.

The Lagrange top is one of the most classical examples of integrable systems. The explicit formulae for the position of the body in space were found by Jacobi \cite{Jacobi2}. For an actual integration of the corresponding equations of motion in terms of elliptic functions
see \cite{ks65} and for a more modern account \cite{gz88,rm82}.

\subsection{Recurrence relations}
According to \cite{gz88} the invariant manifold of the Lagrange top is isomorphic to the invariant manifolds of one-gap solutions of the non-linear Schr\"odinger equation. Their bi-hamiltonian structures may be identified as well.

So, for the Lagrange top there are two known Poisson bivectors $P'$ compatible with the canonical
bivector $P$ (\ref{p0-e3}) \cite{gz88,r82}:
\bq\label{lin-poi}
P'_1=\left(\begin{smallmatrix}
0& x_3& -x_2& 0& 0& 0\\-x_3& 0& x_1& 0& 0& 0\\x_2& -x_1& 0& 0& 0& 0\\0& 0& 0& 0& -\frac{a}2& 0\\0& 0& 0& \frac{a}2& 0& 0\\0& 0& 0& 0& 0& 0
\end{smallmatrix}\right)\qquad\mbox{\rm and}\qquad
P'_2=\left(\begin{smallmatrix}
0& J_3& -J_2& 0& \frac{a}2& 0\\-J_3& 0& J_1& -\frac{a}2& 0& 0\\J_2& -J_1& 0& 0& 0& 0\\0& \frac{a}2& 0& 0& 0& 0\\-\frac{a}2& 0& 0& 0& 0& 0\\0& 0& 0& 0& 0& 0
\end{smallmatrix}\right).
\eq
They are 2-coboundaries $P'_{1,2}=\mathcal L_{X_{1,2}}(P)$ and the corresponding Liouville vector fields $X_{1,2}$ may be obtained from the corresponding vector fields from \cite{ts07c} by using  contraction of $so^*(4)$ to $e^*(3)$.

The Poisson pencil $P_\lambda=P'_{1}-\lambda P$ has one non-trivial polynomial Casimir (\ref{polcas})
\[
H^1(\lambda)=C_1,\qquad
H^2(\lambda)=2\lambda^2C_2+\lambda H_2+aH_1,
\]
while  the second Poisson pencil $P_\lambda=P'_{2}-\lambda P$ has two non-trivial Casimirs
\[
H^1(\lambda)=-2\lambda C_1+H_2,\qquad H^2(\lambda)=-2\lambda C_2+aH_1.
\]
Using the corresponding recurrence relations
\ben
0&=&P'_1dC_1,\nn\\
P'_1H_1=0,\qquad aPdH_1=P'_1dH_2,
\qquad PdH_2&=&2P'_1dC_2,
\en
and
\ben
P'_2dH_2&=0, \qquad PdH_2&=-2P'_2dC_1,\nn \\
P'_2dH_1&=0,\qquad  aPdH_1&=-2P'_2dC_2,
\en
we can easily get integrals of motion $H_{1,2}$ (\ref{int-lag}) starting with the known Casimir functions $C_{1,2}$ (\ref{caz-e3}).

The Poisson tensors $P'_1$ and $P'_2$ are compatible to each other, i.e. $[P'_1,P'_2]=0$. So, existence of the triad $P,P_1',P'_2$ of mutually compatible Poisson bivectors leads to tri-hamiltonian structure for the Lagrange top. Reducing  \`{a} la Marsden--Ratiu this tri-Hamiltonian structure we can get the separated variables for the Lagrange top. The reduction may be not unique, since possibly different separated variables
can be constructed on the symplectic leaf of $P$ \cite{mt02}.

\subsection{Poisson structures having the same foliation by symplectic leaves}

The generic solution of the equation $P_fdC_{1,2}=0$ may be parametrized by
two vector functions $f=(f_1,f_2,f_3)$ and $g=(g_1,g_2,g_3)$
\bq\label{anz}
P^f=\left(\begin{array}{cc}
(x,f)\,x_M&f\otimes (x\wedge J)+Q\\-\left[f\otimes (x\wedge J)+Q\right]^T&
-(x\wedge J)_3\, g_M
\end{array}\right),
\eq
where
\[
Q=\left(\begin{array}{rrr}
~x_2(x\wedge g)_1&~x_2(x\wedge g)_2&~x_2(x\wedge g)_3\\
-x_1(x\wedge g)_1&-x_1(x\wedge g)_2&-x_1(x\wedge g)_3\\
0&0&0
\end{array}\right)\,.
\]
Here $(u\otimes v)_{ij}=u_iv_j$ and $(u\wedge v)_j$ is the $j$-th entry of the crossproduct
$u\wedge v$ of two vectors $u$ and $v$.

For any given integrable system on $e^*(3)$ functions $f$ and $g$ have to satisfy  one algebraic equation
\bq\label{hpf-eq}
\{H_1,H_2\}_f=0
\eq
and  overdetermined system of algebro-differential equations (\ref{m-eq}).
To solve these equations for the Lagrange top we use some hypothesis about the functions
$f$.
\subsubsection{Solution 1}
If we put $f_3=0$ and $(x,f)=0$, then one gets
\[
f_1=\dfrac{-\arccos\left(\dfrac{x_3}{|x|}\right)}{(x\wedge J)_3}\,x_2,\qquad
f_2=\dfrac{-\arccos\left(\dfrac{x_3}{|x|}\right)}{(x\wedge J)_3}\,x_1\,\]
and
\ben
g_1&=&\dfrac{\arctan\left(\dfrac{x_1}{x_2}\right)}{(x\wedge J)_3}\,J_1,\qquad
g_2=\dfrac{\arctan\left(\dfrac{x_1}{x_2}\right)}{(x\wedge J)_3}\,J_2,\nn\\
g_3&=& \dfrac{-\arccos\left(\dfrac{x_3}{|x|}\right)}{(x\wedge J)_3}J_3+\dfrac{x_3\left(\arccos\left(\dfrac{x_3}{|x|}\right)-\arctan\left(\dfrac{x_1}{x_2}\right)\right)(x_1J_1+x_2J_2)}
{(x_1^2+x_2^2)(x\wedge J)_3}.\nn
\en
The corresponding bivector (\ref{anz}) we designate as $P^f_1$.
In this case
\[
F=\left(\begin{smallmatrix}
\arctan\left(\frac{x_1}{x_2}\right)& 0\\
\\
 2\left[\arctan\left(\frac{x_1}{x_2}\right)-\arccos\left(\frac{x_3}{|x|}\right)\right]
 \left(J_3-\frac{x_3(x_1J_2+x_2J_2)}{x_1^2+x_2^2}\right)\qquad
 &\arccos\left(\frac{x_3}{|x|}\right)
\end{smallmatrix}\right)
\]
The eigenvalues of $F$ are the separated variables
\[q_1=\arctan\left(\frac{x_1}{x_2}\right),\qquad q_2=\arccos\left(\frac{x_3}{|x|}\right),\]
which coincide with the Euler angles $\phi$ and $\theta$, respectively. The canonically conjugated momenta read as
\[
p_1=-J_3,\qquad  p_2=-J_1\cos\left(\arctan\left(\frac{x_1}{x_2}\right)\right)
+J_2\sin\left(\arctan\left(\frac{x_1}{x_2}\right)\right).
\]
In variables $(q,p,C)$ two compatible bivectors $P$ and  $P^f_1$ have the standard form
\bq\label{dn-ten}
P=\left(
 \begin{array}{ccc}
 0 & I&0 \\
 -I & 0&0\\
 0&0&0
 \end{array}
 \right)\,,\qquad P^f_1=\left(
 \begin{array}{ccc}
 0 & \mbox{\rm diag}(q_1,q_2)&0 \\
-\mbox{\rm diag}(q_1,q_2) & 0 &0\\
 0&0&0
 \end{array}
 \right)\,.
\eq
Using variables $(q,p,C)$ we can easily prove that projections of the linear bivectors $P'_1$ and $P'_2$ (\ref{lin-poi}) can not be associated with the Euler angles.

\subsubsection{Solution 2:} If we put  $f_3=0$ and $(x,f)\neq0$ then one gets
\bq
\label{l-sol3}\\
f_1= \dfrac{x_1-ix_2}{(x\wedge J)_3}J_2+\dfrac{x_1-ix_2}{(J_1-iJ_2)^2|x|},\qquad
f_2=-\dfrac{x_1-ix_2}{(x\wedge J)_3}J_1-\dfrac{x_1-ix_2}{(J_1-iJ_2)^2|x|}\nn
\eq
and
\[
g_m=-\dfrac{J_1-iJ_2}{(x\wedge J)_3}J_m.
\]
The corresponding bivector (\ref{anz}) we designate as $P^f_2$. In this case
\[
F=\left(\begin{array}{cc}
0\quad& -\dfrac{i}{2(x_2+ix_3)} \\
a\quad& -\dfrac{J_2+iJ_3}{x_2+ix_3}
\end{array}\right)\]
and one gets complex separated variables
\[
q_{1,2}=-\dfrac{J_2+iJ_3\pm \sqrt{ (J_2+iJ_3)^2-2ia(x_2+ix_3) }  }{2(x_2+ix_3)},\qquad i^2=-1.
\]
In variables $(q,p,C)$ two compatible bivectors $P$ and  $P^f_2$ have the form (\ref{f-ten})
\[
P=\left(
 \begin{array}{ccc}
 0 & I&0 \\
 -I & 0&0\\
 0&0&0
 \end{array}
 \right)\,,\qquad P^f_2=\left(
 \begin{array}{ccc}
 0 & -\mbox{\rm diag}(\frac{1}{q_1},\frac{1}{q_2})&0 \\
\mbox{\rm diag}(\frac{1}{q_1},\frac{1}{q_2}) & 0 &0\\
 0&0&0
 \end{array}
 \right)\,,
\]
At the first time these separated variables have been appear in framework of the Sklyanin method and then have been recovered in \cite{mt02} by reduction of the
the compatible linear Poisson bivectors $P'_1$ and $P'_2$ (\ref{lin-poi}). In variables $(q,p,C)$
these linear bivectors look like
\[
P'_1=\left(
 \begin{array}{ccc}
 0 & \mbox{\rm diag}\left(\frac{a}{2q_1},\frac{a}{2q_2}\right)&w_1 \\
-\mbox{\rm diag}\left(\frac{a}{2q_1},\frac{a}{2q_2}\right) & 0 &w_2\\
-w_1&-w_2&0
 \end{array}
 \right)\] and
 \[
 P'_2=\left(
 \begin{array}{ccc}
 0 & -\mbox{\rm diag}\left(\frac{a^2}{4q_1^2},\frac{a^2}{4q_2^2}\right)&w_3 \\
\mbox{\rm diag}\left(\frac{a^2}{4q_1^2},\frac{a^2}{4q_2^2}\right) & 0 &w_4\\
-w_3&-w_4&0
 \end{array}
 \right)\,.
\]
The matrix elements of $w_k$ are brackets $\{q_i,C_j\}'_m$ and $\{p_i,C_j\}'_m$, which are some nontrivial rational functions on the separated variables and Casimirs. For instance
\[
\{q_1,C_2\}'_1=\frac{iq_1p_2}{q_1-q_2}.
\]
We can see that reductions of $P'_1$ and $P^f_2$ on symplectic leaf of $P$ are identical up to multiplication on a constant. Roughly speaking in this case reduction consists of removing the last rows and the last columns of $P'_1$ and $P^f_2$.

\subsubsection{Solution 3:} If  we put $f_1=0$ then one gets $(x,f)=1$ and
\bq\label{l-sol2}
g_m=\dfrac{ax_m}{2(x\wedge J)_3(x_3-ix_2)},\qquad f_2=f_3=\dfrac{1}{x_2+ix_3}\,.
\eq
The corresponding bivector (\ref{anz}) we designate as $P^f_3$.
In this case
\[
F=\left(\begin{array}{cc}
J_1-iJ_2& 0\\
a(x_1-ix_2)+\dfrac{a(x_1-ix_2)^2}{|x|(J_1-iJ_2)^2}\qquad&
-\dfrac{x_1-ix_2}{|x|(J_1-iJ_2)}
\end{array}\right)
\]
and the separated coordinates are
\[
q_1=J_1-iJ_2,\qquad q_2=-\frac{x_1-ix_2}{|x|(J_1-iJ_2)}\,.
\]
In variables $(q,p,C)$ two compatible bivectors $P$ and  $P^f_3$ have the form (\ref{dn-ten}) and, therefore, $P^f_3$ is 2-coboundary. The corresponding separated equations, the Lax matrices, the $r$-matrix formalism and the B\"{a}cklund transformations could be found in \cite{kuz04}.

\vskip0.5truecm
\par\noindent
The Poisson bivectors $P^f_k$, $k=1,2,3$, are incompatible to each other, i.e. $[P^f_i,P^f_j]\neq 0 $ at $i\neq j$. Moreover, they are incompatible with the linear bivectors $P'_{1,2}$ (\ref{lin-poi}) as well. It means that  we have different bi-hamiltonian structures associated with the Lagrange top. This fact deserves further investigation.

\section{Another integrable systems on $e^*(3)$}
\setcounter{equation}{0}
\subsection{The Goryachev-Chaplygin top}
The well-known Goryachev-Chaplygin case  in rigid body dynamics
is described by the following integrals of motion
\bq\label{gor}
H_1=J_1^2+J_2^2+4J_3^2+ax_1\,,\qquad
H_2=2(J_1^2+J_2^2)J_3-ax_3J_1\,, \qquad a\in \mathbb R\,.
\eq
On the fixed level $(x,J)=0$ of the second Casimir function the Hamilton function $H_1$ commutes with an additional cubic integral of motion $H_2$. This fact ensures the integrability of the Goryachev-Chaplygin case.

Substituting anzats (\ref{anz}) into  (\ref{m-sys}) at $f_1=0$ one gets the following solution
\[f_2=0,\qquad f_3=-1,\]
and
\bq\label{g-gch}
g_1= -\dfrac{J_1J_3}{(x\wedge J)_3},\qquad
g_2= -\dfrac{J_2J_3}{(x\wedge J)_3},\qquad
g_3 =\dfrac{J_1^2+J_2^2}{(x\wedge J)_3}
\eq
The corresponding polynomial  $P^f$ (\ref{anz}) has been obtained in \cite{ts07a} by using the $r$-matrix formalism and the Sklyanin brackets.

\begin{rem}
The same bivector $P^f$ could be easier found by using  Liouville vector field $X$ with polynomial entries:
\[
P^f=\mathcal L_X(P_0), \quad X=\sum X_m(z)\partial/\partial z_m, \qquad z=(x,J).
\]
If we suppose that $X_m(z)$ are arbitrary quadratic polynomials on $M$
\bq\label{anz2}
X_m=\sum_{ij}^n c_m^{ij}z_iz_j,\qquad m=2,\ldots,6, \qquad c^m_{ij}\in \mathbb C,
\eq
 then from (\ref{m-sys}) one easily gets
\bq\label{gor-X}
X=\left(\begin{array}{cccccc}0,& x_3J_2,& -x_2J_2,& -J_1J_3,& 0,& -J_2^2-J_3^2\end{array}\right).
\eq
In contrast with rational functions (\ref{g-gch}) here we have simple polynomials only. As a usefull by-product we directly prove that the bivector $P_f$ is 2-coboundary in the corresponding Poisson-Lichnerowicz cohomology.
\end{rem}

In this case the control matrix $F$ is equal to
\bq\label{gor-F}
F=\left(
    \begin{array}{cc}
      2J_3 & -1 \\
      -J_1^2-J_2^2 & 0
    \end{array}
  \right)\,, \qquad
\eq
and  its eigenvalues
\bq\label{gor-var}
q_{1,2}=J_3\pm\sqrt{J_1^2+J_2^2+J_3^2~}
\eq
satisfy to the following dynamical equations
\bq\label{gor-eq}
(-1)^j\,({q}_1-{q}_2)\dot{{q}_j}=2\sqrt{\mathcal P({q}_j)^2-|x|^2a^2{q}_j^2}\,,
\qquad \mathcal P(\lambda)=\lambda^3-\lambda H_1+H_2\,.
\eq
These equations are reduced to the Abel-Jacobi equations and,
therefore, they are solved in quadratures.

\subsection{The Sokolov system on the sphere}
Let us consider another integrable  at $(x,J)=0$ system on $e^*(3)$ \cite{soktmf}
with  integrals of motion second and fourth order:
\bq\label{sok-int}
\begin{array}{l}
{H}_1=J_1^2+J_2^2+2J_3^2+a(x_3J_1-J_3x_1)+2bJ_3,\\
\\
{H}_2=(J_1^2+J_2^2+J_3^2)(2J_3+2b-ax_1)^2\,.
\end{array}
\eq
Using the same anzats (\ref{anz2}) for the components  $X_m(z)$ of the Liouville vector field as for the Goryachev-Chaplygin top one gets the same solution (\ref{gor-X}) of the equations (\ref{m-sys}). In this case the control matrix reads as
\bq\label{sok-F}
{F}=\left(
    \begin{array}{cc}
    J_3& \dfrac{1}{2(2J_3+2b-ax_1)}\\
       2(J_1^2+J_2^2+J_3^2)(2J_3+2b-ax_1)& J_3
    \end{array}
  \right)\,.
\eq
Its eigenvalues coincide with the Chaplygin variables (\ref{gor-var}), which are the separated variables for the Sokolov system too.

\subsection{The Kowalevski top.}
Let us consider the Kowalevski top with
the following integrals of motion
\ben
H_1&=&J_1^2+J_2^2+2J_3^2-2bx_1,\nn\\
\label{kow-top}\\
H_2&=&\Bigl((J_1+iJ_2)^2+2b(x_1+ix_2)\Bigr)\Bigl((J_1-iJ_2)^2+2b(x_1-ix_2)\Bigr).\nn
\en
Solution of the equations (\ref{m-sys}) has been constructed in \cite{ts07b}
by using the $r$-matrix formalism and the reflection equation algebra.
In our notations this solution is defined by
\ben
f_1&=& -2J_1-\frac{(2x_1J_2-x_2J_1)\Bigl(b(x_2J_2+x_3J_3)+J_1J_3^2\Bigr)}{J_2^2(x\wedge J)_3},\nn\\
f_2&=&J_2-\frac{J_1^2+bx_1}{J_2}-
\frac{x_1(2J_3^2-bx_1)}{(x\wedge J)_3}+\frac{x_2J_3(2J_1J_3+bx_3)}{J_2(x\wedge J)_3}
+\frac{x_1(J_1J_3+bx_3)^2}{J_2^2(x\wedge J)_3}
\nn\\
f_3&=&J_3-\frac{b x_2^2J_3}{J_2(x\wedge J)_3}+\dfrac{(J_1J_3+bx_3)(J_1(x\wedge J)_3-bx_1x_2)}{J_2^2(x\wedge J)_3}\nn
\en
and
\ben
g_1&=&\frac{b(x_1J_1+x_3J_3)}{(x\wedge J)_3}+\frac{bx_2(J_1^2-J_3^2+bx_1)}{J_2(x\wedge J)_3}+\frac{J_1(J_1J_3+bx_3)^2}{J_2^2(x\wedge J)_3},
\nn\\
g_2&=& \frac{2bx_2J_1}{(x\wedge J)_3}+\frac{b^2x_2^2+J_1J_3(J_1J_3+bx_3)}{J_2(x\wedge J)_3}+\frac{bx_2J_3(J_1J_3+bx_3)}{J_2^2(x\wedge J)_3},
\nn\\
g_3&=& \frac{bx_3J_1}{(x\wedge J)_3}+\frac{bx_2(J_1J_3+bx_3)}{J_2(x\wedge J)_3}+\frac{J_3(J_1J_3+bx_3)^2}{J_2^2(x\wedge J)_3}\,.\nn
\en
The corresponding separated variables $q_{1,2}$ are the famous Kowalevski variables \cite{ts07b}. In these variables bivectors $P$ and $P^f$ have the form (\ref{dn-ten}). It allows us to prove that the second bivector $P_f$ is the 2-coboundary in the corresponding Poisson-Lichnerowicz cohomology.

\section{Concluding remarks}
The main result in this paper is construction of the different bi-Hamiltonian structures for the  Lagrange top, which have the same foliation by symplectic leaves. The corresponding three incompatible Poisson bivectors may be associated with the 2-coboundaries in the Poisson-Lichnerowicz cohomology defined by canonical bivector $P$ on $e^*(3)$.

As a last remark, we observe that similar bi-hamiltonian structures exist for some other integrable systems on $e^*(3)$, for instance for the Kowalevski top.  The similar  2-coboundaries in the Poisson-Lichnerowicz cohomology on $so^*(4)$ were considered in \cite{ts07c}.

The research was partially supported by
the RFBR grant 06-01-00140.

\end{document}